\documentstyle[editedvolume,epsf]{crckapb} 

\input psfig.tex

\newcommand{\be}{\begin{equation}}
\newcommand{\ee}{\end{equation}}
\def\lsim{\lower.5ex\hbox{$\; \buildrel < \over \sim \;$}}
\def\gsim{\lower.5ex\hbox{$\; \buildrel > \over \sim \;$}}

\begin{opening}
\title{Classification of Light curves of the Black Hole Candidate GRS1915+105}

\author{Anuj Nandi, Sivakumar G. Manickam and Sandip K. Chakrabarti$^1$}
\institute{S. N. Bose National Centre For Basic Sciences\\
	JD Block, Salt Lake, Sector-III, Calcutta-700091, India\\
	email: anuj@boson.bose.res.in, sivman@boson.bose.res.in, chakraba@boson.bose.res.in}
\end{opening}

\begin{document}

\begin{abstract}
{\small The black hole candidate GRS 1915+105 exhibits a rich variety of 
variability. Assuming that the earlier paradigm of black hole accretion
which includes a shock is applicable for this system as well, one can classify these 
variabilities into four classes. We present light curves of these classes
and discuss the basis of our classification.}
\end{abstract}
\smallskip



\runningtitle{Classification of Light Curves of  GRS1915+105}

\noindent $1$ Also, Honorary Scientist at Centre for Space Physics, IA-212, Salt Lake City, Sector-III, Calcutta-700097\\

\noindent Published in Indian Journal of Physics, 74(B), No. 5, P. 331

\section{Introduction}

GRS1915+105 is a stellar mass black hole candidate in our galaxy which exhibits
very rich time-variability. It is termed as a micro-quasar since it has most of the
features of a quasar, such as a superluminal jet which moves at a speed of $98$ percent of the
velocity of light [1-2].  %(Mirabel and Rodriguez, 1994; see, Mirabel 1999). 
Morgan et al. [3] pointed out that the source is sometimes in low-hard state,
but in other times it goes to flare state with considerable variations 
in amplitude and quasi-periodic oscillation frequency. They showed that there are
several time-scales of quasi periodic oscillations (QPO) ranging from 67Hz
to 0.01 Hz (See, also, Muno et al. [4]). Recently, it has been pointed out [5-7] that 
% (Chakrabarti, 1999 IJP; Manickam \& Chakrabarti, 1999; Chakrabarti \& Manickam 2000)
there is a distinct correlation between the oscillation frequency and the duration 
of oscillation. This correlation has been explained by invoking repeated
formation and cooling of the wind generated from the centrifugal pressure dominated boundary
layer (CENBOL) of the black hole [5, 7]. 
(Wind production from accretion flows are also  discussed in Das \& Chakrabarti [8].)
Several types of oscillations have been plotted in Manickam and Chakrabarti [6]. 

Given that accretion disk models around black holes are pretty well understood [9-10], 
and variation of accretion rates, outflow rates, shock locations, cooling processes etc. can make 
specral change dramatically,  it is not perplexing 
as to why GRS1915+105 exhibits such a rich variety of oscillations. 
It was already  pointed out [11-12] that oscillation of shock waves
could be responsible for the QPO. Later, this model was proven to be correct [6,7]  
where it was shown that soft X-rays (0-4 keV) emitted 
from the pre-shock region, do not participate in QPOs and only hard 
X-rays (4-13 keV) show quasi-periodic behaviour.

In this {\it Rapid Communication}, we present a classification
of the light curves based on our understanding of the black hole accretion process according to which
no only Keplerian matter, but also substantial amount of sub-Keplerian matter accretes 
into a black hole and the sub-Keplerian flow often produces stationary or non-stationary shocks.
We find that although there are several types of light curves, the number of classes could be
as few as four. In our picture, the post-shock flow is heated up and puffed up 
and intercepts soft photons from the pre-shock region.
In the pre-shock region, Keplerian disk is situated in the equatorial plane and is flanked by 
poorly emitting sub-Keplerian matter above and below. The post-shock flow
produces winds and outflows, rates of which depend on the spectral
states of the black hole [13]. In the hard states, there is a continuous 
outflow at a smaller rate, while in the soft states, no outflow is produced. 
When the compression ratio of the shock is intermediate, outflow rate 
is maximum and flares and variabilities are prominent.

That the outflows are originated from CENBOL has now been confirmed by several workers.
Mirabel [2] and Dhawan et al. [14] found evidence of IR flares and radio flares which 
are directly correlated with the X-rays produced by inverse-Comptonization at the base of the
jet. Similarly, Fender [15] concluded that the base of the jet IS the same place where 
X-rays are originated.  Junor, Biretta \& Livio [16] found that jets in M87
must be originated within a few tens of Schwarzschild radii.

If our paradigm as described above is correct, and the X-ray variability is primarily due to
variability of the hard X-rays, then it is conceivable that the light curves could be
classified according to the photon counts from pre- and post- shock flows. This is 
precisely what we do. In the next Section, we briefly present the theory of photon
emission from the pre- and post-shock flows. In \S 3, we plot the light curves and
the softness ratios and classify the light curves into four classes. Finally,
in \S 4, we draw our conclusions.

\section {Brief Theory of Photon Emission from an Accretion Disk Around Black Holes}

For a Keplerian disk, the surface flux of soft X-rays emitted is given by [17],
%17 ss84
$$
F= 5 \times 10^{26} (\frac{{\dot M}_{17}}{M^2}) r^{-3} I {\rm erg \ cm^{-2} sec^{-1}}.
\eqno{(1)}
$$
Here, $M$ is the mass of the black hole, measured in units of $M_\odot$, the mass of the
sun, ${\dot M}_{17}$ is the mass accretion rate of the {\it Keplerian component} 
in units of $10^{17}$ gms/sec, $r$ is the distance from the
black holes measured in units of $r_s=GM_{BH}/c^2$ ($M_{BH}$ is the mass of the black hole,
$G$ is the gravitational constant, $c$ is the velocity of light all measured in cgs units), $I=1-(6/r)^{1/2}$.
For an optically thick flow, the black body radiation that is emitted locally is that of temperature,
$$
T_s(r) = [4F(r)/a]^{1/4}= 5 \times 10^7 M^{-1/2} ({\dot M}_{17})^{1/4} r^{-3/4} I^{1/4}   \ \ K
\eqno{(2)}
$$
Here $a$ is the radiation constant.
Close to $r=20$ and for $M=10$, the temperature is roughly $1.4$keV which corresponds to soft-X-ray
radiation. A part of this radiation is intercepted by the post-shock region [18] and is re-emitted
after being energized by inverse-Comptonization process.  If the accretion 
rate is very small, inverse-Comptonization still leaves the post-shock 
region hotter and hard radiation as well as  winds are produced. If both 
the Coulomb heating and inverse-Comptonization are important, the electrons are 
roughly $({m_p/m_e})^{1/2}$ times cooler that the protons: 
$T_e \sim (m_e/m_p)^{1/2} T_p$. In the post-shock region: $T_p \sim 10^{11}$K 
and $T_e \sim 2 \times 10^9$K. Power-law radiation emitted by successive scattering 
can produce radiation starting from $2$ to $200$ keV or more. Even higher 
energy radiations can be emitted by synchrotron processes.

When Keplerian accretion rate is very high, they cool the post-shock electrons 
completely and the shock vanishes. The post-shock region also resembles 
like an optically thick Keplerian flow, but the inner edge of the disk 
($r<6$) produces hard radiation due to bulk-motion Comptonization [18]. 

In the event of oscillating shocks, the winds produced may not leave 
the system and stalled jets are produced.  Matter in the subsonic 
region of the wind falls back to equatorial disk as soon as it is cooled 
due to inverse-Comptonization. This periodicity gives an expression for 
the duration of QPOs which is found to be well correlated with the 
QPO frequency [5, 7]. Various types of QPOs which are observed could 
be due to variations of accretion rates in Keplerian and sub-Keplerian 
components and shock location (which is determined by the specific 
energy and specific angular momentum of the inflow). 

\section {Classification of Light Curves}

Figure 1 shows possible variations of the light curves of GRS1915+105 as observed 
by the RXTE satellite. Twelve panels are marked. Panels 3 and 6 have more than one light curve
(separated by dashed line), as they are similar but  with subtle difference. Along X-axis 
is the time elapsed in 
seconds since the beginning of the observation. Spectral analysis of the 1st 
panel suggests that it is purely in hard state. There is a prominent
QPO whose frequency may change from time to time and photon count number may also change
significantly. Final three panels (10-12) contain 
light curves of those days on which spectral states are soft. There are no QPOs in these
days. Spectral fits indicate high temperature and high photon spectral index. 
The ninth panel contains a light curve where two semi-soft
spectra with different photon counts are seen. Count rate varies very significantly. 
In the remaining seven panels (2-8) photons jump in between two distinct states, one with a low photon count 
(off state) and the other with a high photon count (on state). These twelve types have 
been designated as  $\chi$, $\alpha$, $\nu$, $\beta$, $\lambda$, $\kappa$, 
$\rho$, $\mu$, $\theta$, $\delta$, $\gamma$ and $\phi$ respectively by Belloni
et al. [19]. They also plot hardness ratios of these cases where (B/A) vs. (C/A)
was depicted. Here, A, B, C are photon counts in (0-5) keV, (5-12) keV 
and (12-60) keV ranges. These hardness ratios show various well known features 
such as Atoll, Z-type etc. very similar to what was observed in neutron star candidates.

According to the paradigm of Keplerian/sub-Keplerian flow, the following should be happening 
close to a black hole [18]: Keplerian flow moves in the equatorial plane while sub-Keplerian flow
moves above and below the plane. The Keplerian flow itself becomes sub-Keplerian close to the black hole.
The combined sub-Keplerian flow then continues to move towards the black hole and forms a   
standing or oscillating shock if the specific energy is positive. Otherwise, the combined flow
forms a smaller centrifugal barrier dominated region and passes through the inner sonic point. 
The post-shock region reprocesses soft photons from the Keplerian disk and emits essentially hard
radiation provided Keplerian accretion rate is small enough, otherwise, the post-shock region
also cools down [18]. While hot, the post-shock region produces winds. If the outflow rate is
large enough, it may intercept soft and hard photons and re-process. If the outflow is `failed',
namely, do not pass beyond the sonic point due to various factors such as cooling down
by inverse Comptonization [5, 7], it may fall back onto the post-shock region and  the object
may behave as if the accretion rate had gone up. Most of the on/off transitions can be easily
interpreted in this way.

If the pre-shock flow is indeed the source of the soft photons, photons originating in (0-3) keV range
should be roughly proportional to the accretion rate in the cool component (basically
Keplerian). Thus, photon number may show time variation (due to periodic change in the `accretion
rate'). However, no QPO should be seen. Chakrabarti \& Manickam [7] demonstrated this. 
The harder photons ($E>3\ keV$) would usually come from the post-shock flow. Since spectra intersect at
around $17$keV, and for $E>17$keV, photon number is not large, we make our choice of 
A, B and C to be those in ranges (0-3) keV, (3-17) keV and (17-60) keV  respectively.
According to our paradigm, roughly speaking, A, B and C should be proportional to each other,
(since B and C produced by interception of soft photons. Of course, 
soft X-ray absorption makes matter more complex.)
and whenever hardness or softness ratios are plotted
basically straight lines are expected, instead of Atoll, Banana and Z shapes which 
do not give any insight into the problem.
Figure 2 shows twelve panels (in the same sequence as Fig. 1) where B/C is plotted along X-axis and A/C is 
plotted along the Y-axis. We call this a `softness ratio' diagram.
In some of the panels (No. 2-4, 6-8) the ratio is zoomed in
to show details. One observes that panels (9-12), which are for soft or nearly soft spectra, the plots are
roughly linear and the lower-left end starts at sufficiently large number compared to the
other panels. Power density plots do not show evidence of QPO in these types of light curves.
Light curves of all other panels show QPO. 

The softness diagrams could be classified into the followings: 

\noindent 1. HARD Class (H): Panel 1\\
\noindent 2. SOFT Class (S): Panels 10-12\\
\noindent 3. SEMI-SOFT Class (SS): Panel 9\\
\noindent 4. INTERMEDIATE Class (I): Panels 2-8\\

Each of these  classes have subclasses in terms of variation in light curves
and softness ratios. For instance, in Class H, counts and QPO frequencies vary
and the spot mark in Panel 1 moves around. In Class S, the slopes and the co-ordinates
or the lower-left point varies, but the general nature is similar (Panels 10-12). 
In Class-I, the duration of the off and the on state
may vary systematically from one sub-class to the next. The duration of the on-state (high count)
may be almost zero (Panel 7). At the same time, the co-ordinate of the lower left
corner point in the softness ratio diagram varies systematically. As discussed in Section 2
(Eq. 1) soft photon count A is directly related to the accretion rate in the Keplerian 
disk and hard photon count B is related to the degree of interception,
and sum of Keplerian and sub-Keplerian accretion rate (total matter in the
post-shock region). Thus, different members of this sub-class are directly related to these
accretion rates. The common ground is that the duration of the off-states in all of these sub-classes 
are found to obey the correlation with QPO frequency [7]. The process of generating 
light curves purely from theoretical consideration is more involved (as feedback and other non-linear
processes are present) and would be reported in future.

\section{Concluding Remarks}

We have presented all possible types of light curves and divided them into four
classes according to the nature of light curves and the softness ratios. We find that
the black hole candidate GRS1915+105 stays most of the time in  Class H state
(which has QPO), while touring  around other classes (only off states of 
which have QPOs), depending on the accretion rates in the Keplerian
and sub-Keplerian flows. In Class S, the black hole is in soft state and no QPO
is observed. Our simplified classification is based on the black hole accretion
paradigm that explains the spectral state transitions as well as QPOs. In future,
detailed theoretical modeling would be presented.

\section{Acknowledgments}

SGM acknowledges a grant `Quasi-Periodic Oscillations of Black Holes' 
from ISRO and AN acknowledges a DST grant No. SP/S2/K-14/98 for supporting their Fellowship.

{}

\newpage
\centerline{Figure Captions}

\noindent Fig. 1: All possible light curves of the black hole candidate GRS 1915+105. Along X-axis
is time in seconds and along Y-axis is photon counts in units of $10^3$. In Panels 3 and 6
two days of curves, differing slightly, have been put.

\noindent Fig. 2: Softness Ratio A/C vs. B/C is plotted for all the 12 panels shown in
Fig. 1. A, B, and C are photon counts within intervals (0-3) keV, (3-17) keV and (17-60)
keV respectively. Panel 1 belongs to Class-H, Panel 9 belongs to Class-SS, Panels (10-12)
belong to Class-S and the rest belong to Class-I. See, text for details.

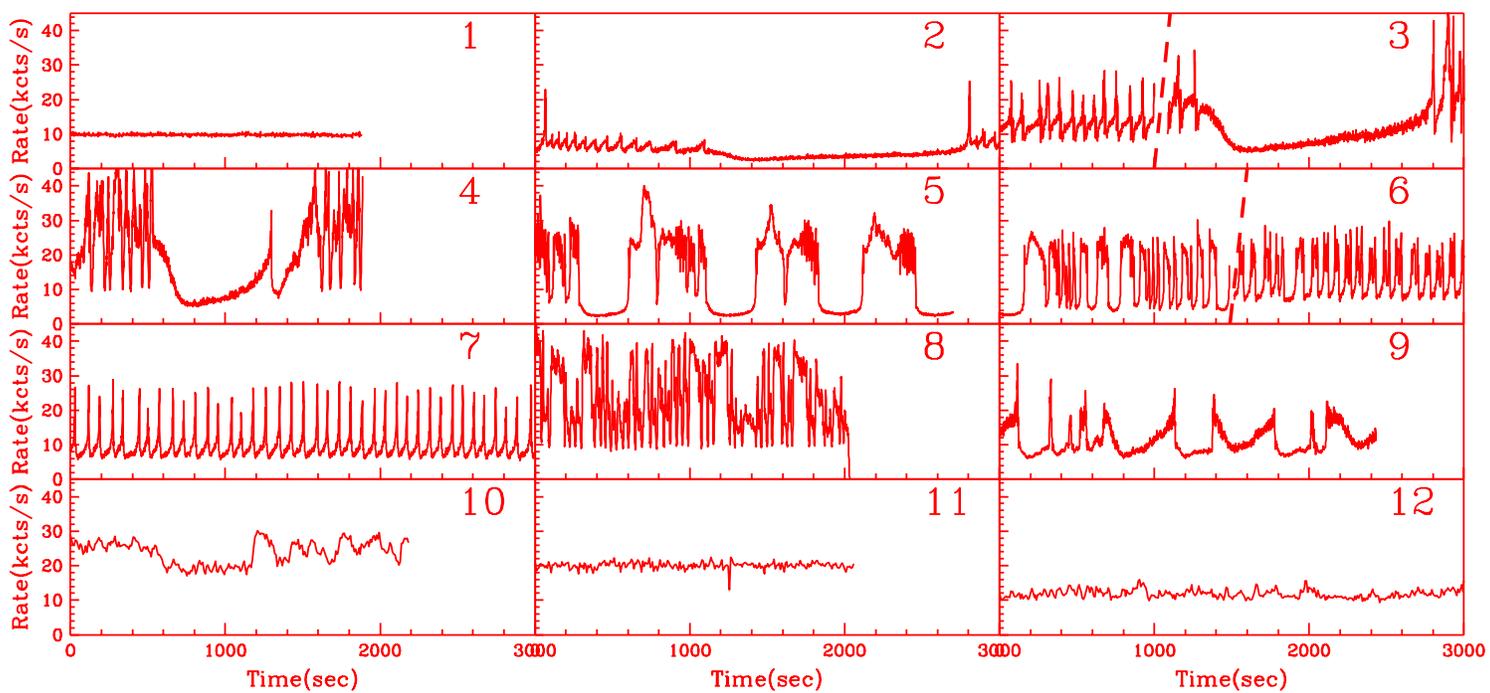
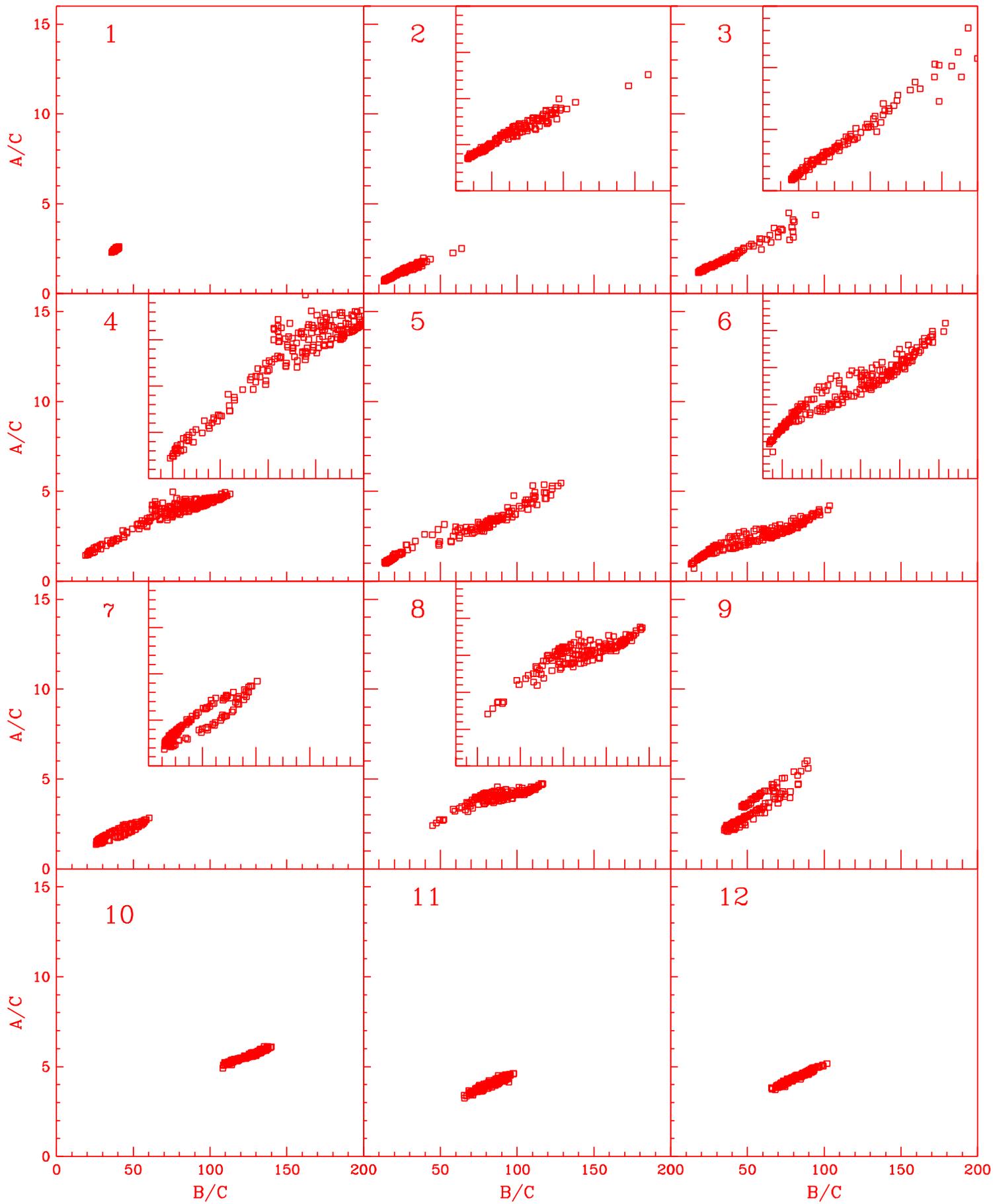
\end{document}